\documentclass[useAMS,usenatbib]{mn2e}
\usepackage[dvips]{graphicx}
\usepackage{txfonts}
\usepackage{amssymb}
\usepackage{indentfirst}
\usepackage{epsfig}

\newcommand{\OVII}{\mbox{{O\,{\sevensize VII}}}}
\newcommand{\OVIII}{\mbox{{O\,{\sevensize VIII}}}}
\newcommand{\OIX}{\mbox{{O\,{\sevensize IX}}}}
\newcommand{\NeIX}{\mbox{{Ne\,{\sevensize IX}}}}
\newcommand{\NeX}{\mbox{{Ne\,{\sevensize X}}}}

\newcommand{\MgXI}{\mbox{{Mg\,{\sevensize XI}}}}
\newcommand{\MgXII}{\mbox{{Mg\,{\sevensize XII}}}}
\newcommand{\MgXIII}{\mbox{{Mg\,{\sevensize XIII}}}}

\newcommand{\SiXIII}{\mbox{{Si\,{\sevensize XIII}}}}

\newcommand{\FeXIX}{\mbox{{Fe\,{\sevensize XIX}}}}
\newcommand{\FeXVII}{\mbox{{Fe\,{\sevensize XVII}}}}

\newcommand{\Lya}{\ensuremath{\hbox{Ly}\alpha~}}
\newcommand{\A}{\AA~}

\def\xmm{{\it XMM-Newton }}

\begin{document}
\title[Charge Exchange X-ray Emission of M82]{
		Charge Exchange X-ray Emission of M82: K$\alpha$ triplets 
			of O\,{\Large\bf VII}, Ne\,{\Large\bf IX}, and Mg\,{\Large\bf XI}} 

\author[J. Liu et al.]{Jiren Liu$^{1,}$\thanks{E-mail: jirenliu@nao.cas.cn},
	Shude Mao$^{1,2}$
 and Daniel Wang$^{3}$\\
$^{1}$National Astronomical Observatories, 20A Datun Road, Beijing 100012, China\\
$^{2}$Jodrell Bank Centre for Astrophysics, University of Manchester,
	  Manchester, M13 9PL, UK \\
$^{3}$Department of Astronomy, University of Massachusetts, Amherst, MA 01002\\
}
\date{}

\maketitle

\begin{abstract}

Starburst galaxies are primary feedback sources of mechanical energy and
metals, which are generally measured from associated
X-ray emission lines assuming that they are from the thermal emission of
the outflowing hot gas. Such line emission, however, can also arise 
from the charge exchange X-ray emission (CXE) between highly ionized ions 
and neutral species. 
To understand the feedback of energy and metals, it is crucial to determine 
the origin of the X-ray emission lines and to distinguish the 
contributions from the CXE and the thermal emission. 
The origin of the lines can be diagnosed
by the K$\alpha$ triplets of He-like ions, because the CXE
favors the inter-combination and forbidden lines, while the thermal emission
favors the resonance line. We analyze the triplets of 
\OVII, \NeIX, and \MgXI\ observed in the \xmm reflection grating spectra 
of the starburst galaxy M82. The flux contribution of the CXE 
is 90\%, 50\%, and 30\% to the \OVII, \NeIX, and \MgXI\ triplet, respectively.
Averaged over all the three triplets, the contribution of the CXE is
$\sim50\%$ of the total observed triplet flux. To correctly understand the 
hot outflow of starburst galaxies, it is necessary to include the CXE.
Based on the measured CXE contributions to the \OVII, \NeIX, and
\MgXI\ triplets, we estimate the relative abundances of O, Ne, and Mg of the 
outflow and find they are similar to the solar ratios.

\end{abstract}

\begin{keywords}
atomic processes -- plasmas -- ISM: jets and outflows -- ISM: abundances --
galaxies: starburst -- galaxies: individual: M82 -- X-rays: ISM
\end{keywords}

\section{Introduction}

Starburst galaxies are the primary sources that eject energy and
metals into the intergalactic medium. The properties of the outflow are 
fundamental to the understanding of the feedback process of galaxies. The energy
and metals of the outflow are generally measured from the X-ray emission, 
particularly emission lines, associated with starburst galaxies. 
In past studies, the X-ray emission lines are assumed to come 
from the hot outflow itself, the temperature and metal 
abundances of which are extracted based on thermal models.

The X-ray line emission, however, can arise not only from the hot
gas, but also from the interaction between the hot gas and the
neutral cool gas. For example, when the solar wind interacts with a comet, 
the highly ionized ions in the wind can readily pick up electrons from 
the neutral species of the 
comet. The product ions remain highly ionized and are left in excited
states. They will emit X-ray photons when they decay to ground states.
This process is called charge exchange X-ray emission 
(CXE, also called charge transfer) 
and explains the bright cometary X-ray emission \citep[e.g.][]{Lis96, Cra97,
	Cra02}.
It can be represented by:
\begin{equation}
A^{q+} + N \rightarrow A^{(q-1)+*} + N^{+},
\end{equation}
where a highly ionized ion $A^{q+}$ (like \OVIII, \NeX) picks up an electron 
from a neutral species $N$ (like H, H$_2$), producing an exited ion $A^{(q-1)+*}$, 
which will emit X-ray photons when it decays to the ground state. 
For the historical studies and recent developments of the CXE, we refer 
to \citet{Den10} and references therein.

Different from the thermal emission, the CXE contributes only emission lines.
If the X-ray line emission, or part of it, is due to the CXE, the
measurement of the thermal and chemical properties of the hot outflow
based on thermal-only models will be misleading.
To correctly understand the hot outflow, it is crucial to reveal the 
origin of the X-ray emission lines
and to distinguish the contributions from the CXE and the thermal emission.

Based on the tight spatial correlation of H$\alpha$ and X-ray emissions,
\citet{Lal04} speculated the importance of the CXE between the hot outflow
and the cool halo gas for starburst galaxies. 
From Suzaku and \xmm observations of the cap above the disk of M82, 
	 \citet{Tsu07} showed 
a marginal detection of an emission line at 0.459 kev, which may be due to
the CXE of CVI ($n=4\rightarrow1$). 
From \xmm EPIC spectra of the central region of M82, \citet{Ran08} also 
reported two lines (around 10 and 16 \AA), which can not be adequately 
accounted for by 
thermal models and may be contributed by the CXE from neutral Mg and \OVIII.

One diagnostic that can be used to determine the origin of the X-ray emission lines
is the K$\alpha$ triplet between $n=2$ shell and $n=1$ ground state of He-like ions,
which consist of a resonance line, two inter-combination lines, and a forbidden 
line \citep[for a recent review, see][]{Por10}.
For a thermal plasma, the electron collisional excitation is efficient and 
favors the resonance line, while for the CXE, the de-excitation favors the triplet
states and thus the inter-combination and forbidden lines. 
Therefore, the line ratios of the triple can determine the origin of the X-ray
emission lines. 
If the forbidden line dominates, the CXE will be the most feasible origin,
because the hot plasma is already highly ionized and photonization is
negligible. 

In this letter we analyze the K$\alpha$ triplets of He-like ions
of \OVII, \NeIX, and \MgXI\ of M82
observed by \xmm Reflection Grating Spectrometers \citep[RGS,][]{den01}. 
M82 is one of the nearby brightest galaxies in X-ray and regarged as 
a prototypical starburst. A bipolar outflow originated from the nuclear 
starburst extends beyond 5$'$ (about 5 kpc) from the disk 
\citep[e.g.,][]{Ste03}.
\citet{Ran08} have analyzed the \xmm RGS \OVII\ triplet of M82
and they fitted three Gaussians to the blended triplet. 
They showed that the line intensity ratios are marginally consistent with the CXE.
In this letter, we decompose the triplets into the CXE and the thermal emission
using the laboratory measured CXE spectrum of \OVII\ triplet. 
As a result, we can quantify the contribution of the CXE to the observed 
K$\alpha$ triplets of He-like ions, including \OVII, \NeIX, and \MgXI. 
The RGS also allow us to study the \OVII\ triplet at different spatial
regions.

The local emissivity of the CXE of one particular line
(assuming single electron capture)
$P_{c}\propto n_{A^{q+}}V\sigma n_{neu}$, 
where $n_{A^{q+}}$, $n_{neu}$, V, and $\sigma$ denote the ion density,
the neutral density, the relative velocity, and the charge exchange
cross section, respectively. 
If we know the CXE fluxes of the \OVII, \NeIX, and \MgXI\ triplets, 
we can estimate the relative abundance of O, Ne, and Mg with the 
 temperature information. It provides important insight 
to the composition of the hot outflow, especially when the assumption 
of the thermal-only origin of the X-ray line emission is poor.
We will make a first attempt to 
estimate the relative abundances of O, Ne, and Mg using the CXE fluxes
of their He-like triplets.

The paper is structured as follows. We describe the observation data 
in \S 2. The analysis results are presented in
\S 3. Our conclusion and discussion are given in \S 4.
 Errors are given at 1 $\sigma$ confidence level.

\section{Observation data }

We use four archival \xmm RGS observations of M82, the obs-IDs of which are
0112290201, 0206080101, 0560590201, and 0560590301. 
The total effective exposure time is $\sim$ 120 ks after removing intense 
flare periods, which doubles the useful dataset used by \citet{Ran08}.
The most recent version of Science Analysis System (SAS, 10.0) is used for
the reduction of photon events.
In these observations, the RGS slit is generally not parallel to 
the outflow direction and the extracted emission lines are broadened 
due to the tilt of the outflow. Thus we limit our extraction
to events within 30$''$ radius of the cross-dispersion direction.
We also extract the \OVII\ triplet from $-90''$ to $-30''$
and from $30''$ to $60''$ of the cross-dispersion direction, where
the \OVII\ triplet is bright enough for further study.
For the region larger than $60''$,
the \OVII\ triplet is heavily blended by the structures of the X-ray emission
and does not allow a simple fitting as we do below, so it is excluded.

There are two RGS on the \xmm. CCD4 in RGS2 covering
the \OVII\ triplet failed, and CCD7 in RGS1 covering the \NeIX\ triplet
also failed. Thus, the \OVII\ triplet data are from RGS1 and 
the \NeIX\ triplet data from RGS2, while the \MgXI\ data are from 
both RGS1 and RGS2. The \SiXIII\ triplet
around 6.7 \A is neglected here as the effective area is
small and the photon counts are too few to obtain robust
results. 

\section{Results}

\begin{figure}
\includegraphics[height=2.3in]{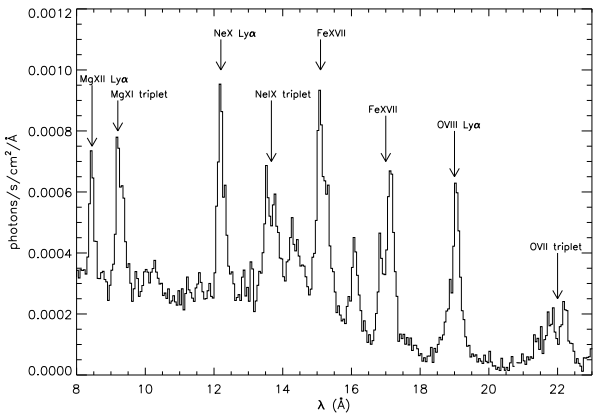}
\caption{Fluxed \xmm RGS spectrum of M82 extracted from 
the central 30$''$ radius of the cross-dispersion direction.
Both RGS1 and RGS2 are combined together.
The spectrum is dominated by \Lya lines of H-like and He-like 
states of O, Ne, and Mg and neon-like Fe lines.
The triplets of \OVII, \NeIX, and \MgXI\ are analyzed below.
}
\end{figure}

The observed \xmm RGS spectrum of M82, corrected for the effective area,
is shown in Figure 1. As can be seen, the emission is dominated by 
\Lya lines of H-like and He-like states of O, Ne, and Mg and 
neon-like Fe lines. Below we analyze the
K$\alpha$ triplets of He-like ions of \OVII, \NeIX, and \MgXI. The 
intensity ratio of resonance, inter-combination, and forbidden lines 
can tell the origin of the X-ray emission lines.

\subsection{\OVII\ triplet}
\begin{figure*}
\includegraphics[height=2.3in]{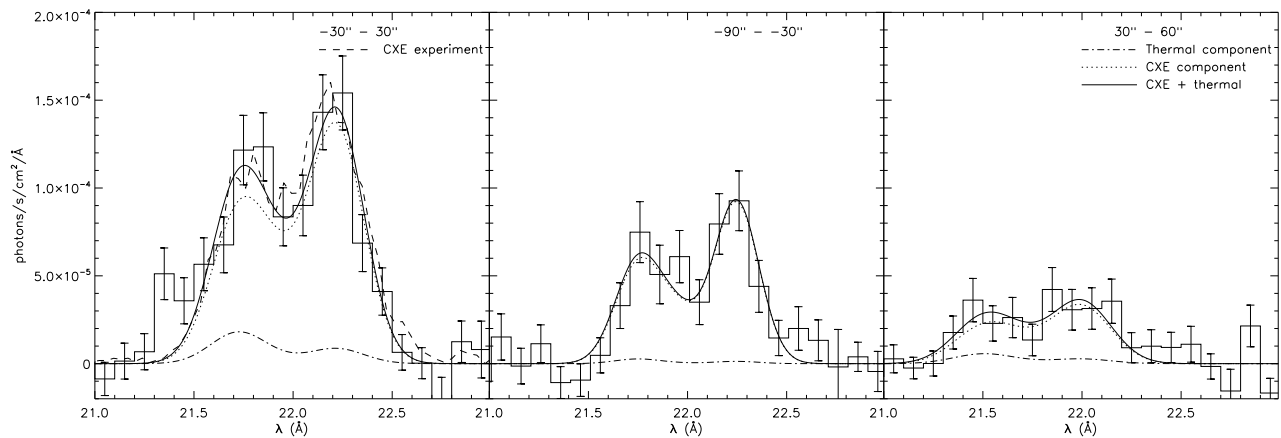}
\caption{RGS spectra of the \OVII\ triplet in three extraction regions.
A continuum, which is calculated as
the interpolation between the line wings, has been subtracted. 
Laboratory data for the CXE of the \OVII\ triplet 
by \citet{Bei03} is over-plotted as the dashed line in the left panel.
The RGS spectrum agrees well with the CXE data.
The solid line is the sum of the fitted model of the CXE (dotted line) and the 
thermal emission 
(dot-dashed lines). Most of the flux of \OVII\ triplet is due to the CXE.} 
\end{figure*}

Let's first focus on the observed \OVII\ triplet within the 
central $-30''$ -- $30''$ region,
which is plotted in the left panel of Figure 2.
A continuum interpolated between the line wings 
has been subtracted. It shows clearly that the 
forbidden line is stronger than the resonance line.
The dominance of the forbidden line can not result from the thermal
excitation, but is a signature of the CXE.
The laboratory measured CXE of \OVII\ triplet using the spare X-ray
micro-calorimeter from ASTRO-E (Suzaku) mission \citep{Bei03} 
is over-plotted as dashed line in Figure 2. The laboratory measured
CXE of \OVII\ triplet fits the observed \OVII\ triplet well.

\begin{table*}
\caption{Results of fitting the CXE and thermal emission to 
	the triplets of \OVII, \MgXI, and \NeIX.} 
\begin{tabular}{c|c|c|c|c|c|c|c|cccc}
\hline
 &$f_c(10^{-5})^a$&$f_t(10^{-5})^a$&
 $\Delta\lambda$(\AA)&
 $\sigma_{\lambda}$(\AA)&$\chi^2$/dof&r(\AA)&i(\AA)&f(\AA)&$t_r$&$t_i$&$t_f$  \\
\hline
\OVII($-30''-\: 30''$)& 8.6$\pm$1.3 & 1.0$\pm$1.2 & 0.12$\pm$0.02& 0.14$\pm$0.01 &7.0/6
&21.6&21.8&22.1&0.624&0.070&0.306\\ \hline
\OVII($-90''-\: -30''$)& 4.9$\pm$0.9 & 0.1$\pm$0.9 & 0.15$\pm$0.01& 0.11$\pm$0.01 &6.4/6
&21.6&21.8&22.1&0.624&0.070&0.306\\ \hline
\OVII($30''-\: 60''$)& 2.3$\pm$0.8 & 0.3$\pm$0.8 & -0.1$\pm$0.05& 0.15$\pm$0.03 &5.2/6
&21.6&21.8&22.1&0.624&0.070&0.306\\ \hline
\MgXI &2.6$\pm$ 1.4 & 7.1$\pm$1.5 & 0.03$\pm$0.006&0.05$\pm$0.005& 11.1/9
&9.17&9.23&9.31&0.587&0.101&0.312\\ \hline
\NeIX(${\rm \frac{Ne}{Fe}}=1.5^b$) & 5.1$\pm$1.2&5.7$\pm$1.0$^c$ & 0.06$\pm$0.007&
0.06$\pm$0.006& 10.2/8 &13.45&13.55&13.7&0.138&0.027&0.076\\ \hline
\NeIX(${\rm \frac{Ne}{Fe}}=4^b$) & 4.6$\pm$1.4&6.5$\pm$1.2$^c$ & 0.06$\pm$0.008
&0.06$\pm$0.007& 11.5/8 &13.45&13.55&13.7&0.262&0.052&0.144\\ \hline

\end{tabular}
\begin{description}
\begin{footnotesize}
\item
Note: for the meaning of symbols, see eq. (2); 
$^a$$f_c$ and $f_t$ are in units of photons/s/cm$^2$;
$^b$${\rm \frac{Ne}{Fe}}$ is in units of the solar ratio; 
$^c$the value $f_t$ of \NeIX\ is for the thermal emission of both
 \NeIX\ triplet and \FeXIX\ lines.
\end{footnotesize}
\end{description}
\end{table*}

To quantify the contributions of the CXE and the thermal emission, we fit the 
observed 
triplet with two components of the CXE and the thermal emission. The CXE component
consists of three Gaussians centered on the triplet lines,
 the ratios of which are fitted 
to the laboratory measured CXE triplet by Beiersdorfer et al. (2003). 
The thermal component also consists of three Gaussians, the ratios of which are
calculated at $T=8\times10^6$ K, which is the estimated temperature
of the hot gas (see \S 3.2). 
The modeled spectrum can be written as  
\begin{equation}
f_{model}=\frac{1}{\sqrt{2\pi}\sigma_{\lambda}}\sum_{j=r,i,f}
(f_cc_j+f_tt_j)\exp\left[-\frac{(\lambda-\lambda_j-\Delta\lambda)^2}
{2\sigma_{\lambda}^2} \right], 
\end{equation}
where $\lambda_{j}$ is the wavelength of resonance, inter-combination, and
forbidden lines, $c_{j}$ the corresponding 
ratios for the CXE,  $t_j$ the ratios for the thermal emission, 
$f_c$ and $f_t$ is the flux of the CXE and the thermal emission respectively.
The CXE ratios $c_{r,i,f}$=0.317, 0.149, 0.535;
the thermal ratios $t_{r,i,f}$ are calculated 
using APEC 2.0 \citep{Apec01} and listed in Table 1.
We assume that the dispersion $\sigma_{\lambda}$ and the wavelength 
shift $\Delta\lambda$ are the same for both components.
Since the observed RGS lines are dominated by
the spatial extent of the source, it means that we assume both components 
are from a similar spatial volume.
The fitted results are listed in Table 1 and
plotted in Figure 2. 
The thermal component only accounts for 10$\pm12$\% of the
observed \OVII\ triplet. 
As the CXE dominates, the results are insensitive to the adopted temperature.
For example, if we adopte $T=5\times10^6$ K, the thermal fraction is 11$\pm14$\%.
For higher temperatures, the thermal ratios $t_{r,i,f}$ are almost unchanged
and the results are not affected. Thus we conclude that most of the observed \OVII\
triplet is due to the CXE.

We can estimate the CXE contribution to the observed \OVIII\ \Lya(19 \AA) flux
within the same region,
which is $\sim1.8\times10^{-4}$ photons/s/cm$^2$. 
In \S 3.2, we will see that the flux ratio of \MgXII\ \Lya to 
\MgXI\ triplet
indicates a temperature around $8\times10^6$ K.
At such a high temperature, the ion fraction of \OVIII\ is negligible 
and \OIX\ ions dominate charge exchange collisions. 
While it is possible for an \OIX\ ion to capture two electrons to become 
an \OVII\ ion directly, the cross section of double-electron charge exchange is 
generally lower
than 20\% of that of single-electron charge exchange \citep{Gre01}.
Thus we consider the case of multiple charge exchange collisions.
That is, an \OIX\ ion experiences one charge exchange collision becoming 
an \OVIII\ ion, then becomes an \OVII\ ion through another collision.
As the transition ratio of
$n(\ge3\rightarrow1)$ to $n(=2\rightarrow1)$ is measured to be similar for
the collision of \OIX\ and \OVIII\ \citep{Gre01}, in the case of 
multiple collisions,
the emission of an \OVII-triplet photon is coupled with the 
emission of an \OVIII-\Lya photon. In other words,
the CXE flux of \OVIII\ \Lya is the same as that of \OVII\ triplet. 
Since the fitted CXE flux of \OVII\ triplet 
is $8.6\times10^{-5}$ photons/s/cm$^2$,
we expect that about half of the total observed flux of \OVIII\ \Lya is due
to the CXE. The CXE contribution to the \OVIII\ \Lya emission will be larger if the
assumption of multiple collisions is not fulfilled.

The observed \OVII\ triplets from $-90''$ to $-30''$
and from $30''$ to $60''$ are plotted in the middle and right panels respectively.
They also show the dominance of the forbidden line. 
That is, the CXE occurs not only at the starburst nucleus ($\sim30''$), 
but also at the outflow regions.  The fitted results are listed in Table 1.
The different $\Delta \lambda$ reflects the tilt of the outflow relative 
to the cross-dispersion direction mentioned in 
\S 2. The observed \OVIII\ \Lya fluxes of the two regions are 8.6 and 
$4.9\times10^{-5}$ photons/s/cm$^2$ respectively. The observed flux 
ratios of \OVII\ triplet to \OVIII\ \Lya are similar for all three
regions, which implies a same scenario of multiple charge exchange
collisions.

\subsection{\MgXI\ triplet}

\begin{figure}
\includegraphics[height=2.3in]{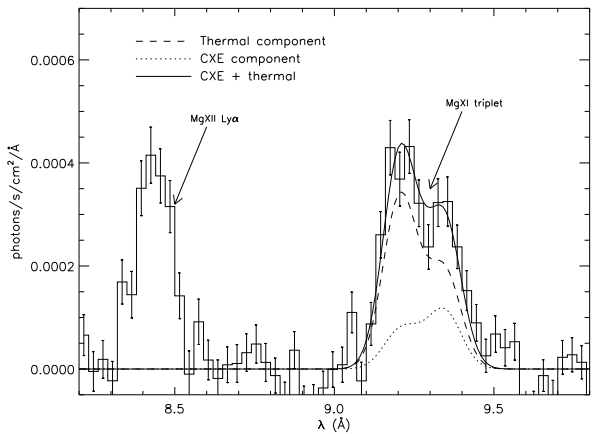}
\caption{RGS spectrum of the \MgXI\ triplet. An interpolated continuum
has been subtracted. The solid line is
the sum of the CXE component (dotted line) and the thermal 
component (dashed line). The \MgXII\ \Lya line is also plotted.}
\end{figure}

The observed \MgXI\ triplet and \MgXII\ \Lya line are plotted in Figure 3.
Different from the \OVII\ triplet, the \MgXI\ triplet is dominated by 
the resonance line, i.e., by the thermal emission.
This enables an estimation of the
temperature of the hot gas assuming there is no CXE. The ratio (0.57) of 
the observed flux of \MgXII\ \Lya line ($5.6\times10^{-5}$ photons/s/cm$^2$)
to that of \MgXI\ triplet ($9.7\times10^{-5}$ photons/s/cm$^2$) implies 
a temperature of $8\times10^6$ K.
We fit the observed \MgXI\ triplet with both 
the CXE and the thermal emission.
The ratios $c_{r,i,f}$ of the CXE are assumed to be the 
same as those of the laboratory measured CXE of the \OVII\ triplet. 
The ratios of the thermal
component are calculated at $T=8\times10^6$ K.
The fitted results are plotted in Figure 3 and listed in Table 1. 
The contribution of CXE is about $30\pm15$\% to the observed \MgXI\ triplet.

We can perform a self-consistency check on the temperature estimation 
by taking into account the CXE.
The ion fraction of \MgXIII\ and \MgXII\ at $T=8\times10^6$ K 
is 0.28 and 0.44 respectively.
In the case of multiple charge exchange collisions, the CXE flux of 
\MgXII\ \Lya line is 40\% of that of \MgXI\ triplet. 
Excluding the CXE flux, the remaining thermal flux of \MgXII\ \Lya  
is $4.6\times10^{-5}$ photons/s/cm$^2$. 
The ratio of the thermal flux of \MgXII\ \Lya to that of \MgXI\
triplet ($7.1\times10^{-5}$ photons/s/cm$^2$) is 0.65, which 
corresponds to a temperature of $\sim8.2\times10^6$ K.
If single collision happens, the thermal flux of \MgXII\ \Lya is
$4\times10^{-5}$ photons/s/cm$^2$,
corresponding to a temperature of $\sim8\times10^6$ K. 
Thus the temperature estimated without the CXE is consistent with
that considering the CXE.

\subsection{\NeIX\ triplet}

\begin{figure}
\includegraphics[height=2.3in]{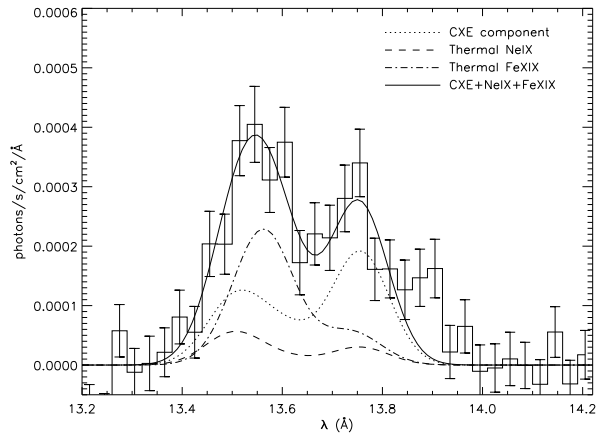}
\caption{RGS spectrum of the \NeIX\ triplet. 
The continuum has been subtracted. The \NeIX\ triplet is mixed with \FeXIX\ lines.
The solid line is
the sum of the CXE (dotted line), the thermal \NeIX\ triplet (dashed line),
and the thermal \FeXIX\ emission (dot-dashed line).
}
\end{figure}

The observed \NeIX\ triplet is plotted in Figure 4.
It is similar to the \MgXI\ triplet (right part of Fig.3). 
The \NeIX\ triplet, however, is mixed 
with \FeXIX\ lines. At $T=8\times10^6$ K, the \FeXIX\ lines 
are comparable to, or even stronger than, the \NeIX\ triplet. 
We take into account the \FeXIX\ lines 
by including several other 
Gaussians centered on strong \FeXIX\ lines. 
The abundance ratio of
Ne/Fe, obtained by \citet{Ran08} from fitting the RGS spectrum assuming
thermal-only models, is 1.5 times the solar value \citep[calibrated to][]{Lod03}. 
However, their fitted Mg/Fe ratio is 4 times the solar value.
Since Fe is mainly produced in SNe Ia and $\alpha$-elements are enriched 
by SNe II, in principle, $\alpha$/Fe should be similar.
Thus we adopt two values (1.5 and 4) for the Ne/Fe ratio. 
The fitted results are listed in Table 1 and
the result of the Ne/Fe ratio of 1.5 is plotted in Figure 4.

We see from Table 1 that the fluxes of the CXE do not change much for
the Ne/Fe ratios of 1.5 and 4. This is because the mixing of \FeXIX\ lines
is mainly around the resonance line, the Ne/Fe ratio does have much effect on
the relative importance of the resonance and forbidden lines. 
The CXE flux of the \NeIX\ triplet ($\sim5\times10^{-5}$ photons/s/cm$^2$) 
dominates over the thermal 
flux of the \NeIX\ triplet (1.4 and 3$\times10^{-5}$ photons/s/cm$^2$ 
for Ne/Fe=1.5 and 4 respectively) and accounts for $\sim50\%$ of the
total observed flux.

\section{Conclusion and discussion}

We have analyzed the K$\alpha$ triplets of \OVII, \NeIX, and
\MgXI\ of the \xmm RGS spectra of M82.
We show that most of the observed \OVII\ triplet is due to the CXE. 
The contribution of the CXE
is 30\% to the \MgXI\ triplet.  
Taking into account the mixing of \FeXIX\ lines,  
the fitted CXE of \NeIX\ triplet dominates over its thermal emission.
Averaged over all the three triplets, the contribution of the CXE is
about 50\% to the total observed flux. 

As stated in the introduction, the local emissivity of the CXE of 
one particular line $P_{c}=(\sum_{q}f_qn_{A^{q+}})V\sigma n_{neu}$,
where the sum is over all possible ionization states that
can charge exchange to produce the line with an efficiency $f_q$.
We can estimate the relative abundances of O, Ne, and Mg based on the
CXE fluxes of their K$\alpha$ triplets. 
We consider the case of multiple charge exchange collisions as discussed in 
\S 3.1.
An additional complication is the foreground absorption. While the column density
$N_H$ can be obtained from fitting the whole spectrum as discussed below,
here we adopt a column density $N_H=0.2\times10^{22}$cm$^{-2}$, which is 
taken from \citet{Ran08}.
Then the absorption factor ($f_{abs}$) relative to the
\OVII\ triplet is 3 and 4.5 for the \NeIX\ and 
\MgXII\ triplet respectively.

We use the relation of the charge exchange cross section $\sigma\propto Z$,
where $Z$ is the element number \citep[e.g.][]{War08}.
We assume the relative velocity is the same for all three elements
as their element numbers are similar.
We also assume the efficiency of producing the triplet $f_q$ is the 
same for all three elements.
The observed CXE flux of the triplet
$f_{c}\propto n_Af_{A}Zf_{abs}$, 
where $n_A$ is the number density for element A and $f_{A}$  the 
fraction of both fully stripped and H-like ions for element A.
This leads to $n_A\propto {f_{c}}/({f_{A}Zf_{abs}})$.
Using the fitted $f_c$ in Table 1, we find that the chemical abundance 
relative to O is 0.16 and 0.06 for Ne and Mg respectively.
These ratios are similar to the solar ratios of 0.15 and 0.07, calculated 
from the solar abundance table of \citet{Lod03}.

The O abundance of the hot gas of M82
is found to be about 4 times smaller than other $\alpha$ elements in
thermal-only models \citep{Read02,Ori04}, which is difficult to explain.
However, the abundance ratios of O, Ne, and Mg estimated from the CXE are 
similar to the solar ratios. Therefore, the low abundance of O obtained 
from thermal-only models is likely to be a consequence of neglecting
the CXE.

As the average contribution of the CXE
to the triplets is $\sim50\%$ and the contribution differs for different lines, 
the CXE will affect the
estimation of the thermal and chemical properties of the hot gas.
In previous studies of M82 \citep[e.g.,][]{Read02,Ran08}, 
multiple-temperature models have been used to fit the observed spectra.
The multiple temperatures are likely due to the neglect of the CXE.
For example, if neglecting the CXE, the temperature estimated from the 
observed flux ratio of \OVIII\ \Lya to \OVII\ triplet will be 
$3.5\times10^6$ K, which is misleading. 
To properly estimate the temperature and metal abundances, it is necessary to fit the 
whole observed spectrum with both the CXE and the thermal emission.
For such a task, a detailed model of the spectra of the CXE,
including the CXE of \FeXVII\, is needed. 
We plan to return to this in a later work.

\section*{Acknowledgements}

We thank Peter Beiersdorfer for kindly providing us the data 
of laboratory measured CXE of the
\OVII\ triplet data, Jifeng Liu for helpful discussions, and our referee
for valuable comments.
This research has made use of \xmm archival data.
XMM-Newton is an ESA science mission with instruments
and contributions directly funded by ESA Member States and the USA (NASA).

\bibliographystyle{mn2e}

\end{document}